\documentclass[aps,amsmath,11pt]{revtex4}
\usepackage{verbatim}
\def\>{\rangle}
\def\su2{$SU(2)$}

\begin{document}

\vspace*{-1.5cm}

\hfill
{NSF-KITP-06-28}\\

\title{A schematic model of neutrinos }
\author{Pavel Kovtun}
\email{kovtun@kitp.ucsb.edu}
\author{A. Zee}
\email{zee@kitp.ucsb.edu}
\affiliation{KITP, University of California, Santa Barbara, CA 93106, USA}%
\date{April 2006}

\begin{abstract}
\noindent  
We show that the observed pattern for neutrino mixing arises naturally 
if neutrinos are composites of more elementary constituents carrying an
$SU(2)$ quantum number we call lepospin.
\end{abstract}

\maketitle

\section{Introduction}

\noindent
The discovery of neutrino oscillations \cite{PDG} 
is the first indication of physics beyond the Standard Model 
of particle physics. 
In addition to quark and charged lepton masses, as well as
mixing in the quark sector, we now have to understand neutrino masses,
as well as mixing in the lepton sector. In this note we will focus on the
lepton sector, and adopt the view that the physics responsible for
mixing is independent of the physics which controls the values of the
masses, much like the separation of the angular and radial parts of the
Schrodinger equation. We assume three Majorana neutrinos. 
The mixing pattern originally proposed by Harrison, Perkins, and Scott
\cite {HPS}

\begin{equation}
\left( 
\begin{array}{l}
\nu_{e} \\ 
\nu_{\mu } \\ 
\nu_{\tau }
\end{array}
\right) =\left( 
\begin{array}{ccc}
-\sqrt{2/3} & \sqrt{1/3} & 0 \\ 
\sqrt{1/6} & \sqrt{1/3} & \sqrt{1/2} \\ 
\sqrt{1/6} & \sqrt{1/3} & -\sqrt{1/2}
\end{array}
\right) \left( 
\begin{array}{l}
\nu_{1} \\ 
\nu_{2} \\ 
\nu_{3}
\end{array}
\right) 
\label{eq:HPS}
\end{equation}
(in the basis where the charged lepton mass matrix is diagonal)
is in excellent agreement with current experimental data. 
In the standard parametrization, 
this mixing pattern corresponds to
$\sin^2(\theta_{13})=0$,
$\sin^{2}(\theta_{23})=1/2$,
$\sin^{2}(\theta _{12})=1/3$.
The simple form of the mixing matrix $V$ defined by (\ref{eq:HPS})
suggests an underlying symmetry.

In this note, after briefly reviewing recent attempts to derive
the above mixing matrix using the traditional approach
\cite{family-symmetry} of postulating
a family symmetry group $G$ which acts on the
three generations of leptons, we suggest
a more speculative approach in which the neutrinos
are regarded as composites. 

In the family symmetry approach, $G$ constrains the form 
of the mass terms. 
Spontaneous breaking of $G$ 
leads to a pattern of fermion masses and mixing. 
One choice for $G$, particularly attractive for
trying to derive $V$, is the non-abelian discrete group $A_4$, 
the symmetry group of the tetrahedron 
\cite{Ma}.
The tetrahedral group has four irreducible representations, 
${\bf 1}$, ${\bf 1'}$, ${\bf 1''}$, and ${\bf 3}$, 
indicated by their dimensions. 

While a number of specific models have been proposed
\cite{Ma, A4-bunch}, 
a general effective field theory analysis
was given recently in \cite{hep-ph/0508278}.
The result of this model-independent analysis could be stated as follows:
under two assumptions, 
{\it (i)} all Higgs fields are weak doublets, and 
{\it (ii)} $A_4$ is broken at the electroweak scale, 
one could obtain a one-parameter family of mixing matrices 
which contains the matrix $V$ in (\ref{eq:HPS}).
To set the relevant parameter to zero so as to obtain $V$ requires
fine tuning.
One could interpret this result to indicate
that deviations from $V$, in particular,
a non-zero $V_{e3}$, should be searched for experimentally. 
Theoretically, any models that satisfy the stated assumptions
of the analysis, no matter how complicated or contrived,
must not be able to arrive at $V$ without fine tuning.
It is, however, easy to break the two assumptions stated above. 

In particular, one can study a seesaw model
containing three right-handed neutrinos
[$SU(2)\times U(1)$ singlets] $N$ in ${\bf 3}$ of $A_4$.
The right-handed neutrinos $N$ acquire mass via a bare mass term $NN$
and a term $\sigma NN$ involving a $SU(2)\times U(1)$ singlet
Higgs $\sigma$ transforming as the ${\bf 3}$ of $A_4$.
The three left-handed lepton doublets $\psi$ are assigned
to ${\bf 3}$ of $A_4$,
while the three right-handed lepton singlets $\psi^c$ are in
${\bf 1}$, ${\bf 1'}$, and ${\bf 1''}$.
The charged leptons acquire mass by the usual Higgs coupling
$\varphi\psi\psi^c$, where the weak doublet $\varphi$
is in ${\bf 3}$ of $A_4$.
As usual in the seesaw mechanism, we couple $\psi$ to $N$
with a Higgs doublet $\chi$, transforming as a singlet under $A_4$.
Apparently, the matrix $V$ could be obtained in this model
if the vacuum expectation values of the three Higgs doublets $\varphi$
are all equal and if the vacuum expectation value of the $\sigma$
point in the 2-direction within $A_4$ \cite{KZ}.
The same conclusion was reached independently in an interesting
paper by He, Keum, and Volkas \cite{hep-ph/0601001}. 
Evidently, we have broken both of the assumptions stated above:
$A_4$ is now broken by the $SU(2)\times U(1)$ singlet Higgs field $\sigma$
at the right-handed neutrino mass scale.
Thus, it appeared that the mixing matrix $V$ can be derived
without fine-tuning.

Unfortunately, the equality of the vacuum expectation values
of the three Higgs doublets $\varphi$ can only be guaranteed 
by $A_4$, but $A_4$ is already broken 
by $\sigma$ at a higher mass scale. 
A picturesque way of saying this is that one can not prevent  
$\sigma$ from talking to $\varphi$ --- the fine-tuning has crept
into the Higgs potential.
He {\it et al.} referred to this difficulty of separating
$\varphi$ and $\sigma$ as the ``sequestering problem''
which they solved by introducing low-energy supersymmetry
\cite{hep-ph/0601001}. 
 
\section{Composite neutrinos}

Given this background, we propose here a rather speculative 
alternative route to deriving $V$. 
Curiously, the required mixing pattern arises naturally 
if one assumes a composite structure for neutrinos. 
We propose that the three left-handed neutrino fields 
$\nu_e$,  $\nu_\mu$, and $\nu_\tau$ are composites of two
constituent fields $\omega$ and $\Omega$, 
each transforming under the triplet representation of an internal 
\su2 group, which we call ``lepospin''.
Physical fields or states are required to have eigenvalue $L_z=0$. 

The crucial observation is that the matrix (\ref{eq:HPS}) is
(up to phase redefinitions) 
exactly the matrix of Clebsch-Gordan coefficients 
when two $L=1$ representations are combined 
to form three states with $L_z=0$. 
(This observation has also been made independently by
Bjorken, Harrison and Scott \cite{BHS}
in passing in their recent paper on a phenomenological analysis
of neutrino mixing.)  
We write, in a slight abuse of notation,
\begin{equation}
{\setlength\arraycolsep{1pt} 
\begin{array}{lcccccc}
|\nu_e\> = |1,0\> |1,0\> & = & \sqrt{\frac23}\,|20\> & + & 0\,|10\> & - & 
\sqrt{\frac13}\,|00\> \\ 
|\nu_\mu\> = |1,1\> |1,-1\> & = & \sqrt{\frac16}\,|20\> & + & \sqrt{\frac12}%
\,|10\> & + & \sqrt{\frac13}\,|00\> \\ 
|\nu_\tau\> = |1,-1\> |1,1\> & = & \sqrt{\frac16}\,|20\> & - & \sqrt{\frac12}%
\,|10\> & + & \sqrt{\frac13}\,|00\>
\end{array}
}  \label{eq:CG}
\end{equation}
Thus mass eigenstates are identified as
$\nu_1=|20\>$, $\nu_2=|00\>$, $\nu_3=|10\>$, 
while the weak eigenstates are composites of
$\omega$ and $\Omega$ with different projection of lepospin.
For example, the left-handed neutrino $\nu_\mu$
is assumed to be a composite of the $L_z=1$ component 
of $\omega$ and the $L_z=-1$ component of $\Omega$.

Remarkably, this line of thought implies the observed mixing matrix $V$. 
Clearly, many questions present themselves which we are unable to answer.  
It is presumably premature at this point to speculate 
on the precise dynamical origin of this apparent compositeness. 
We will restrict ourselves to some remarks. 

Somehow the Hamiltonian projects out the $L_z=0$ states as low energy states. 
One could imagine the Hamiltonian containing
a term like $AL_z^2$ with $A$ large. 
Are the $L_z\neq0$ partners of the neutrinos merely very massive 
and thus will be seen in future accelerators 
or are they not in the physical spectrum?
This is perhaps not too dissimilar from the confinement of quarks. 
In Gell-Mann's original paper on quarks \cite{Gell-Mann}
he supposed that quarks are extremely massive.

As is well-known, current experiments only measure 
the two mass squared differences $m_3^2-m_2^2$ and $m_2^2-m_1^2$, 
and not the masses themselves. 
If indeed neutrino masses follow the ``normal hierarchy''
so that $m_3^2\gg m_2^2\gtrsim m_1^2$ then curiously the $L=2$ state
is the lightest and the $L=1$ state is the heaviest.
So we imagine the effective Hamiltonian to contain
$f(L)+AL_z^2$ with $f(1)\gg f(0)\gtrsim f(2)$.

The matrix $V$ is of course defined in the basis in which
the charge lepton mass matrix is diagonalized. 
Indeed, this in part accounts for the difficulty in deriving $V$
since it results from a ``mismatch'' between the two rotations
diagonalizing the neutrino mass matrix and the charged lepton mass matrix.
In the present composite scheme we have to ``explain''
why the left-handed charged leptons are mass diagonal
in the basis $|1,0\> |1,0\>$, $|1,1\> |1,-1\>$, and $|1,-1\> |1,1\>$
rather than in the states with definite total $L$.
At this point, we could only speculate. 

Long ago, it was noted \cite{Wilczek-Zee-so10}
that the group theory of $SO(10)$ appeared to suggest
a composite structure for quarks and leptons.
Within the spinorial 16-dimensional representation of $SO(10)$,
each quark and lepton field can be coded by a binary string.
For example, the left-handed neutrino corresponds to $|{-}{+}{-}{-}{-}\>$
while the left-handed charged lepton corresponds to $|{+}{-}{-}{-}{-}\>$.
This suggests a composite of five ``spin'' $1/2$ objects
although the authors of \cite{Wilczek-Zee-so10} were ultimately unable
to find a dynamical scheme that worked.
Piling speculation upon speculation, we are tempted to suggest here
that the three left-handed neutrinos correspond to 
$|{-}{+}{-}{-}{-}\>\otimes|1,0\>|1,0\>$, 
$|{-}{+}{-}{-}{-}\>\otimes|1,1\>|1,-1\>$, and 
$|{-}{+}{-}{-}{-}\>\otimes|1,-1\>|1,1\>$,
and similarly for the three charged leptons.
Denote the binary string by $|s_1, s_2,s_3,s_4,s_5\>$,
with $s_i=\pm1$.
As explained in \cite{Wilczek-Zee-so10}, the $W$ boson, for example, 
simply flips $s_1$ and $s_2$ simultaneously,
leaving $s_3$, $s_4$, and $s_5$ untouched.

Within this framework, we may be able to explain
that neutrino and charged lepton mass eigenstates
are rotated relative to each other by coupling
$L_{\omega}$ and $L_{\Omega}$ to $s_1$ and $s_2$.
For instance, the term $f(L)$ in the Hamiltonian
might be multiplied by $(s_1-s_2-2)^2$ and thus
is operative only for the neutrinos.
We imagine that the Hamiltonian contains many terms,
with the essential feature that linkage between the
lepospin sector and the $SO(10)$ sector is such that
for the neutrinos it picks out mass eigenstates
of the total lepospin
${\bf L}={\bf L}_\omega + {\bf L}_\Omega$.

Needless to say, we do not have a detailed dynamical model
any more than the authors of \cite{Wilczek-Zee-so10}.
In particular, it is not clear how to incorporate
the left-handed charged anti-leptons, which correspond to
$|{-}{-}{+}{+}{+}\>\otimes|1,0\>|1,0\>$,
$|{-}{-}{+}{+}{+}\>\otimes|1,1\>|1,-1\>$, and
$|{-}{-}{+}{+}{+}\>\otimes|1,-1\>|1,1\>$.
Similarly with the left-handed anti-neutrinos
whose ``binary string'' is given by $|{+}{+}{+}{+}{+}\>$
[they are the ``odd man out'' in each generation
in the language of $SO(10)$].

The $SO(10)$ structure is meant to be illustrative.
The point is that we need some additional structure
to tell us that for the neutrinos the mass eigenstates
have definite ${\bf L}={\bf L}_\omega + {\bf L}_\Omega$,
but not for the charged leptons.

Knowing almost nothing about the dynamical mechanism
responsible for the binding of $\omega$ and $\Omega$,
we could hardly comment on their spins.
For example, one may be a boson, and another a fermion,
in which case supersymmetry could be relevant
to understanding the binding mechanism.
Alternatively, both $\omega$ and $\Omega$ may be bosons;
it has been known for a long time that fermionic
bound states can appear as composites of bosonic pion fields \cite{Witten}.
If lepospin is realized as a global symmetry at low energies,
then it may be unbroken [in which case one may speculate about
its relation to the custodial \su2], or spontaneously broken
[in which case some of the constituents may be Goldstone bosons,
bound to form massive neutrino states].
If lepospin is a gauge symmetry, then it may be invisible,
much like color is invisible in the low-energy description
of mesons and baryons, or it may be realized in a
strongly coupled Higgs phase.
In any case, a suitable dynamical mechanism of binding should explain why
only states with $L_z=0$ are observed.

Perhaps $\omega$ and $\Omega$ could be produced, 
at the LHC or in cosmic rays. 
The discovery of the composite character of neutrinos
would certainly be exciting.

Of course, our discussion of composite neutrinos leaves many loose ends, 
but if history provides a reliable guide, 
one may also be cautioned against premature judgments. 
For example, before the neutron was discovered,
the nucleus was thought to be made of protons and electrons. 
Obviously, this produces many puzzling questions such as 
why some electrons are bound strongly inside the nucleus 
while others orbit the nucleus. 
There are also difficulties with the spin-statistics of atomic nuclei.
For a more recent example, at the time of Gell-Mann's
landmark paper on quarks \cite{Gell-Mann}, the concept of color,
let alone quantum chromodynamics, was unknown.
As we now know, many of the criticisms leveled at Gell-Mann's paper,
while perfectly reasonable and unanswerable within the physics
of the time, turned out to be misguided.

{\it Acknowledgments.---}
This work was supported in part by the
National Science Foundation under Grant No. PHY99-07949.

\end{document}